# Effect of vibronic relaxation in fluorescence resonance energy transfer: An exact analytical solution

Sangita Mondal[‡], Sayantan Mondal[‡], Kazuhiko Seki[#,]* and Biman Bagchi[‡,]*

[‡]Solid State and Structural Chemistry Unit, Indian Institute of Science, Bengaluru, India
[#]National Institute of Advanced Industrial Science and Technology (AIST), Tsukuba, Japan

*Corresponding authors' emails: bbagchi@iisc.ac.in; k-seki@aist.go.jp

## Abstract

Fluorescence resonance energy transfer (FRET) is widely used as a '*spectroscopic ruler*' to measure fluctuations in macromolecules because of the strong dependence of the rate on the separation (R) between donor (D) and acceptor (A). However, the well-known Förster rate expression that predicts an $R^{-6}$ dependence, is limited by several approximations. Notable among them is the neglect of the vibronic relaxation in the reactant (donor) and product (acceptor) manifolds. Vibronic relaxation can play an important role when the energy transfer rate is faster than the vibronic relaxation rate. Under such conditions, donor to acceptor energy transfer can occur from the excited vibronic states. This phenomenon is not captured by the usual formulation based on the overlap of donor emission and acceptor absorption spectra. Here, we attempt to eliminate this lacuna, by allowing *relaxation in the vibronic energy levels* and adopting a relaxation model to account for vibronic cascading down in the donor manifold. We develop a Green's function based generalized formalism and provide an exact solution for the excited state population relaxation and the rate of energy transfer in the presence of vibronic relaxation. We find and verify that the neglect of vibronic relaxations can significantly alter the energy transfer rate and overestimates the distance between D and A.

## I. INTRODUCTION

Resonance energy transfer is a photophysical process where an optically excited donor molecule transfers its excitation energy to an acceptor molecule. Although this process occurs in a non-radiative manner through Coulombic coupling between the donor and the acceptor, it is often referred to as fluorescence resonance energy transfer (FRET). In the recent past, FRET[1] has become a useful technique, widely employed to study the structure and dynamics of large molecules such as polymers, proteins, DNA, etc. FRET plays a crucial role as a *spectroscopic ruler* because the rate ($k_{DA}$) depends strongly on the distance of separation (R) between the donor (D) and acceptor (A) [Eq.(1)].[2,3] Hence, FRET can be used to measure fluctuating distances in macromolecules, for example, to study protein folding. The celebrated expression of FRET derived by Förster is shown in Eq. (1).[1,4]

$$k_{DA} = k_D \left( \frac{R_0}{R} \right)^6 \tag{1}$$

Here $k_D$ is the rate constant of the radiative decay process of the donor, and $R_0$ is the characteristic Förster radius at which the energy transfer efficiency between the D-A system decreases by 50%. FRET captures the dynamics through optical signatures from D and A. Förster energy transfer rate can be expressed in terms of spectral overlap integral between donor emission spectra [$F_D(\lambda)$] and acceptor absorption spectra [$\varepsilon_A(\lambda)$]. By using Fermi's golden rule (FGR) Eq. (1) can be written as

$$k_{DA} = \frac{9000\kappa^2 ln10}{128\pi^5 N_A \tau_D \eta^4 R^6} \int d\lambda F_D(\lambda) \varepsilon_A(\lambda) \lambda^4 \tag{2}$$

where *J* is defined by

$$J = \int d\lambda F_D(\lambda) \varepsilon_A(\lambda) \lambda^4 \tag{3}$$

Here $J$ denotes the spectral overlap integral, $\kappa^2$ is the orientation factor that depends on the relative orientation of donor emission dipole and acceptor absorption dipole, $N_A$ is Avogadro's number, $\eta$ is the refractive index of the medium, $\tau_D$ is the excited state lifetime of the donor in the absence of an acceptor, $\lambda$ is the wavelength, $F_D(\lambda)$ denotes the normalized steady-state fluorescence spectrum of the donor, and $\varepsilon_A(\lambda)$ is the wavelength-dependent steady-state extinction coefficient of the acceptor.[4–6] There are several experimental, theoretical, and computer simulation studies that investigated FRET and its various aspects.[7–16,17] Although the Förster equation is widely used in different branches of science, the breakdown of the rate equation has been observed both theoretically and experimentally.[6,18–24]

The present work bears several similarities with the theory of electron transfer. The celebrated Marcus theory of electron transfer reactions assumes a small overlap between the electronic orbitals of the reactants in the activated complex.[25] Several reactions in the *Marcus inverted regime* show an enhancement in the rate than those predicted by the Marcus theory.[26] Jortner and Bixon explained these observations in terms of the participation of high-frequency quantum modes.[27] Barbara *et al.* proposed another model where electron transfer consists of solvent polarization mode, a low frequency classical vibrational mode, and a high-frequency vibrational mode.[28] Further investigations have shown that ultrafast polar solvation modes and intramolecular vibrational modes might invoke complex dynamical behavior in photo-induced electron transfer reactions of large molecules.[29,30]

Vibrational states can mediate energy transfer between the electronic states of two molecules. Therefore, before reaching the steady emitting state, the donor can undergo vibronic energy relaxation. This relaxation process does not affect the excitation energy transfer rate if the relaxation rate is much faster than the energy transfer. However, when the energy transfer rate is comparable with the relaxation rate, energy transfer occurs from the non-equilibrium donor state.

The significance of non-equilibrium effects on excitation energy transfer was first characterized by Tekhver and Khizhnyakov, who termed the process as the *hot transfer*.[31] Jortner and co-workers experimentally studied the vibrational-level dependence of excitation energy transfer for various systems in solid and liquid phases.[32] Later Sumi and co-workers generalized the Förster formula in the dynamical regime. They developed a theoretical formalism of energy transfer between two localized states in the presence of vibronic relaxation.[33] They extensively studied excitation energy transfer in complex systems, namely, aggregated chromophores.[21,22]

Recently, conceptual advances have been made by considering the influence of intramolecular vibrational modes. Several studies have been carried out to understand the excitation energy transfer process in the non-equilibrium regime.[34,35] Seibt *et al.* derived the equation of motion for the reduced density matrix of a molecular system with multiple energy transfer channels. They solved the quantum master equation with the help of the projection operator technique and arrived at a generalized expression for the population.[36] Balevičius *et al.* have modeled multi-steps vibrational relaxation of carotenoids and solved it using quantum relaxation theory.[37] Abramavicius and co-workers have studied the role of molecular vibration in excitation dynamics and charge separation. More details on the involvement of vibrational relaxations in carotenoids can be found in the literature.[38–40]

Non-equilibrium generalization of Förster-Dexter (FD) theory was developed for a non-stationary initial state with the harmonic oscillator bath model.[1,41] They investigated the effect of non-equilibrium kinetics on the excitation-bath coupling strength.[42] Their study revealed that the reaction rate shows weak sensitivity on spectral overlap between the stationary emission and absorption spectra in the short-time regime because of non-equilibrium effects. Further generalization of FRET has been achieved for the inelastic condition with quantum mechanical modulation of donor-acceptor coupling in multi-chromophoric systems.[43] Silbey *et al.* developed an extension of the FD theory by considering small polaron transformation to study exciton transport in molecular crystals.[44]

We note that the Förster expression uses a steady-state spectrum that can be measured experimentally and is expressed in terms of oscillator strengths. However, energy transfer can occur during the excited state population relaxation subsequent to the excitation of the donor ground state. Such non-equilibrium effects are also ignored in Förster's expression. To address this scenario, we adopt a phenomenological approach and try to proceed analytically as far as possible with certain limitations. We consider a model by incorporating non-equilibrium effects with some constraints (**Figure 1**). We arrive at closed-form but highly coupled equations and provide the detailed derivation and analyses of these equations. The advantage of our study is, through a straightforward model, we able to capture the non-equilibrium scenario of the energy transfer process, also we explored how it affects the distance measurement.

Here we find that there might exist a range of situations where the use of Förster expression can lead to significantly different value of R. In some cases, the value of R can be smaller or larger in reality, than that predicted by the Förster theory. Therefore, careful scrutiny of the excitation wavelength dependence of the Förster Energy transfer could provide highly relevant information because the rate can depend strongly on the initial state of the vibronic donor manifold.

We organize the rest of the paper as follows. In **section II**, we detail the theoretical formalism and describe the model system chosen in our study. **Section III** contains the generalized solution of the theoretical formalism. **Section IV** consists of the results followed by discussions. Finally, in **section V** we draw general conclusions based on our results.

## II. THEORETICAL FORMULATION

We consider a donor-acceptor (D-A) system where the donor is in the first electronic excited state and the acceptor is in the ground state **(Figure 1).** This state is represented as $|D^*u'; Aw\rangle$, $u'$ and $w$ are the indices of the vibronic energy level of the corresponding electronic state. This initial energy of the system is $(E_{D^*u'} + E_{Aw})$. After the completion of the energy transfer processes between donor and acceptor, the acceptor goes to an excited state (*A\**) and, the donor returns to its ground electronic state. Thus the state of the system is written as $|Du; A^*w'\rangle$ associated with energy $(E_{Du} + E_{A^*w'})$. We denote the energy difference between the excited and ground electronic state of the donor as $E = E_{D^*u'} - E_{Du}$. Energy transfer occurs when $E_{A^*w'} - E_{Aw} = E$.

Let $k_{DA}(E_{D^*u'}, E_{Aw}, E | R)$ be the rate constant associated with energy transfer from $u'$<sup>th</sup> vibronic level of D* to $w$<sup>th</sup> vibronic level of A. Therefore, the total energy transfer rate can be written as $k_{DA}(E_{D^*u'}, E_{Aw}, R) = \sum_E k_{DA}(E_{D^*u'}, E_{Aw}, E | R)$ assuming that all the energy transfer processes are uncorrelated among each other.

We note that in this treatment, the mutual orientation of the donor and acceptor dipoles is ignored. We assume that the energy transfer rates, for all $u'$<sup>th</sup> levels of the donor to the corresponding $w$<sup>th</sup> vibronic energy level of the acceptor, are equal. Such a description has been adapted in order to obtain a closed-form Green's function as the solution for the time-dependent populations. Naturally, this model fails to capture the contribution from the optically dark states but does well in describing the effects of vibronic relaxations.

The population of the $u'^{th}$ vibronic level of the excited state donor and the $w'^{th}$ vibronic level of the ground state acceptor at a time $t$ is denoted by $P_D(D^*u', Aw, t)$. Thus total excited state donor population can be written as the summation of $P_D$ over $u'$ and $w$.

$$P_D(t) = \sum_{D^*u'} \sum_{Aw} P_D(D^*u', Aw, t) \tag{4}$$

Similarly, for the acceptor, one can write

$$P_A(t) = \sum_{Du} \sum_{A^*w'} P_A(Du, A^*w', t) \tag{5}$$

There are several other relaxation processes, namely, vibronic relaxation among the vibrational states of the donor and the $Aw$ vibrational states of the acceptor, which occur simultaneously. Because of these, the excited state population of the donor changes. Hence, we introduce an operator L$_{DA}$ that accounts for vibronic relaxation. According to Kasha's rule, fluorescence occurs only from the lowest vibronic energy level of the first electronic excited state[45] and the rate constant associated with the fluorescence of the donor is denoted by $k_D$.

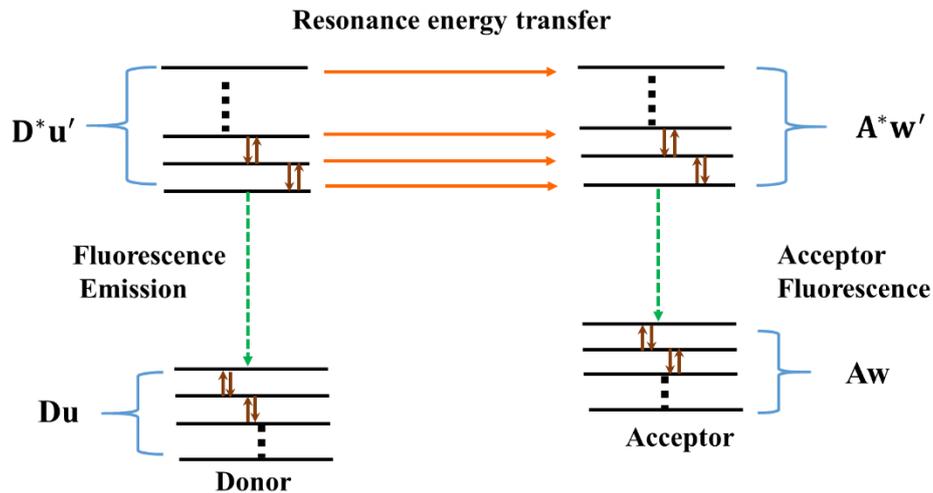

**Figure 1. Schematic representation to show the excitation energy transfer between donor and acceptor. In this model, an infinite number of equispaced vibronic levels of both donor and acceptor are considered. Here u and u′ denote the vibronic sub-levels of the electronic ground state (D) and in the electronic excited state ($D^*$) of the donor respectively. Similarly, for acceptor, vibronic sub-levels of the electronic ground state (A) and electronic excited state ($A^*$) are denoted by indices w and w′ respectively. In this model, it is considered that at an arbitrary time t=0 excitation of donor ($Du \to D^*u'$) has taken place,**

while the acceptor remains in its ground state $Aw$. By virtue of energy transfer (shown by orange arrows), vibronic relaxation (brown arrows), and radiative emission (dashed arrow) the excited state population relaxes. In the acceptor manifold, a similar relaxation mechanism is assumed. The orange arrows represent the energy transfer from donor to acceptor; additionally, we consider the rate of the transfer process is equal for all vibronic sub-level of the excited donor.

Time evolution equation of $P_D(E_{D^*u'}, E_{Aw}, t)$ is given by,

$$\frac{\partial P_D(E_{D^*u'}, E_{Aw}, t)}{\partial t} = L_{DA} P_D(E_{D^*u'}, E_{Aw}, t) - k_{DA}(E_{D^*u'}, E_{Aw} | R) P_D(E_{D^*u'}, E_{Aw}, t) - k_D \delta_{u',0} P_D(E_{D^*u'}, E_{Aw}, t) \quad (6)$$

We apply summation over the entire vibronic manifold to obtain

$$\sum_{D^*u'} \sum_{Aw} \frac{\partial P_D(E_{D^*u'}, E_{Aw}, t)}{\partial t} = \sum_{D^*u'} \sum_{Aw} \left[ L_{DA} P_D(E_{D^*u'}, E_{Aw}, t) - k_{DA}(E_{D^*u'}, E_{Aw} | R) P_D(E_{D^*u'}, E_{Aw}, t) - k_D \delta_{u',0} P_D(E_{D^*u'}, E_{Aw}, t) \right] \quad (7)$$

where the natural boundary conditions of the operator $L_{DA}$ imply[44,45],

$$\sum_{D^*u'} \sum_{Aw} \left[ L_{DA} P_D(E_{D^*u'}, E_{Aw}, t) \right] = 0 \quad (8)$$

By using Eq. (5) and Eq. (8) we obtain

$$\frac{\partial P_D(t)}{\partial t} = -\sum_{D^*u'} \sum_{Aw} \left[ k_{DA}(E_{D^*u'}, E_{Aw} | R) P_D(E_{D^*u'}, E_{Aw}, t) \right] - \sum_{Aw} k_D P_D(E_{D^*u'}, E_{Aw}, t) \quad (9)$$

We assume that the acceptor ground state is in thermal equilibrium. Thus $P_D(E_{D^*0}, E_{Aw}, t)$ can be written in terms of equilibrium ground state population of the acceptor $\rho_{eq}(E_{Aw})$ as

$$P_D(E_{D^*0}, E_{Aw}, t) = P_D(E_{D^*0}, t) \rho_{eq}(E_{Aw}) \quad (10)$$

In the regime of steady state approximation, we can write

$$L_A \rho_{eq}(E_{Aw}) = 0 \quad (11)$$

for the operator $L_A$ in

$$L_{D*A} = (L_{D*} + L_A) \tag{12}$$

By using Eqs. (10) and (11) one can write,

$$L_{D*A} P_D(E_{D*0}, E_{Aw}, t) = (L_{D*} + L_A) P_D(E_{D*0}, t) \rho_{eq}(E_{Aw}) = (L_{D*}) P_D(E_{D*0}, t) \rho_{eq}(E_{Aw}) \tag{13}$$

Also, the rate of energy transfer in terms of the equilibrium ground state population of the acceptor is given by

$$k_{DA}(E_{D*u'} | R) = \sum_{Aw} \rho_{eq}(E_{Aw}) k_{DA}(E_{D*u'}, E_{Aw} | R) \tag{14}$$

As the equilibrium population in the acceptor ground state is normalized, one can write

$$\frac{\partial}{\partial t} \sum_{Aw} P_D(E_{D*u'}, E_{Aw}, t) = \frac{\partial}{\partial t} P_D(E_{D*u'}, t) \sum_{Aw} \rho_{eq}(E_{Aw}) = \frac{\partial}{\partial t} P_D(E_{D*u'}, t) \tag{15}$$

We use Eq.(6) to transform the left-hand side of Eq.(15)

$$\frac{\partial}{\partial t} \sum_{Aw} P_D(E_{D*u'}, E_{Aw}, t) = \sum_{Aw} \frac{\partial}{\partial t} P_D(E_{D*u'}, E_{Aw}, t)$$
$$= \sum_{Aw} \left[ L_{DA} P_D(E_{D*u'}, E_{Aw}, t) - k_{DA}(E_{D*u'}, E_{Aw} | R) P_D(E_{D*u'}, E_{Aw}, t) - k_D \delta_{u',0} P_D(E_{D*u'}, E_{Aw}, t) \right]$$

$$\tag{16}$$

From Eq.(13), (14) and (15)

$$\frac{\partial}{\partial t} P_D(E_{D*u'}, t) = -L_{D*} P_D(E_{D*u'}, t) - k_{DA}(E_{D*u'} | R) P_D(E_{D*u'}, t) - k_D \delta_{u',0} P_D(E_{D*u'}, t) \tag{17}$$

Here $k_{DA}(E_{D*u'} | R) = \sum_E \sum_{Aw} \left[ k_{DA}(E_{D*u'}, E_{Aw}, E | R) \right] \rho_{eq}(E_{Aw})$.

The expression of $k_{DA}(E_{D*u'} | R)$ indicates the sum of all possible energy transfer rates from the state denoted by $D*u'$. By using the Green's function for the operator $L_{D*}$ and by considering the natural boundary condition at the lowest excited state of the donor, the solution can be expressed in the following way in Laplace space;

$$\tilde{P}_D(E_{D^*u'}, S) = \sum_{D^*u'_i} G_{D^*}(E_{D^*u'} | E_{D^*u'_i}, S) P_D(E_{D^*u'_i}, t=0)$$
$$-\sum_{D^*u'_i} G_{D^*}(E_{D^*u'} | E_{D^*u'_i}, S) k_{DA}(E_{D^*u_i} | R) \tilde{P}_D(E_{D^*u'_i}, S) - G_{D^*}(E_{D^*u'} | E_{D^*0}, S) k_D \tilde{P}_D(E_{D^*0}, S) \quad (18)$$

After rearrangements and algebraic manipulations, Eq.(18) can be written as Eq.(19)

$$\sum_{D^*u'_i} \left( \delta_{D^*u', D^*u'_i} + G_{D^*}(E_{D^*u'} | E_{D^*u'_i}, S) k_{DA}(E_{D^*u_i} | R) + G_{D^*}(E_{D^*u'} | E_{D^*0}, S) k_D \delta_{D^*0, D^*u'_i} \right) \tilde{P}_D(E_{D^*u'}, S)$$
$$= \sum_{D^*u'_i} G_{D^*}(E_{D^*u'} | E_{D^*u'_i}, S) P_D(E_{D^*u'_i}, t=0) \quad (19)$$

## III. GENERALIZED SOLUTION OF ABOVE FORMULATION

Here we solve Eq.(19) to obtain the temporal evolution of the donor excited states. In this section, we provide a detailed solution of the formulation provided in **section II**. We follow a generalized procedure to solve the system of differential equations. First, we define the probability generating function as

$$F(z,t) = \sum z^n P_n(t) \quad (20)$$

In Eq.(20), z is the auxiliary variable. To obtain the generating function for the harmonic oscillator (HO), we consider another model system (**Figure 2**) analogous to the *step-Ladder model* which is an efficient tool to study vibronic relaxation of the HO.[46]

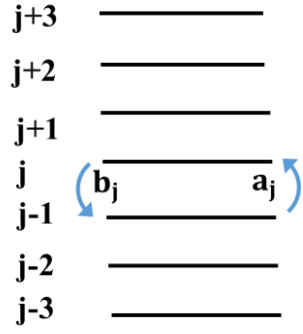

**Figure 2.** A schematic representation of the step-ladder model for the harmonic oscillator. Here j denotes vibration level. $a_j$ is the transition rate from $(j-1)^{th}$ to $j^{th}$ the vibrational levels, similarly $b_j$ is the transition rate that, being at $j^{th}$ level, a jump occurs to $(j-1)^{th}$ level. $a_j$ and $b_j$ are related by $a/b = \exp(-\hbar\omega/k_B T)$.

Thus for linear one-step process and by considering natural boundary condition, we can write the probability generating function as [47]

$$F(z,t) = \frac{(a-b)\left[a(1-\varepsilon)+(a\varepsilon-b)z\right]^m}{\left[a-b\varepsilon-b(1-\varepsilon)z\right]^{m+1}} \quad (21)$$

Where $\varepsilon = e^{(b-a)t}$, $\dfrac{a}{b} = e^{-\theta}$, $\theta = \dfrac{\hbar\omega}{k_B T}$. Now we employ the following identity[48] [Eq.(22)].

$$\frac{(1-t)^m}{(1-t+xt)^{m+1}} = \sum_{n=0}^{\infty} {}_2F_1(-n, m+1; 1; x) t^n \quad (22)$$

In Eq.(22) ${}_2F_1(a,b;c;x)$ denotes a Gaussian hypergeometric function. Therefore, by using Eq.(22) generating function can be written in terms of the hypergeometric function as follows.

$$F(z,t) = \frac{(a-b)\left[a(1-\varepsilon)+(a\varepsilon-b)z\right]^m}{\left[a-b\varepsilon-b(1-\varepsilon)z\right]^{m+1}} \quad (23)$$

$$= \frac{(a-b)}{(a-bz)} \frac{\left[\dfrac{a-bz+\varepsilon(az-a)}{(a-bz)}\right]^m}{\left[\dfrac{a-bz-b\varepsilon+b\varepsilon z}{(a-bz)}\right]^{m+1}}$$

$$= \frac{(a-b)}{(a-bz)} \frac{\left[1-\dfrac{\varepsilon a(1-z)}{(a-bz)}\right]^m}{\left[1-\dfrac{\varepsilon(1-z)}{(a-bz)}\left(1+\dfrac{b-a}{a}\right)a\right]^{m+1}}$$

Therefore, we find

$$F(z,t) = \frac{(a-b)}{(a-bz)} \frac{\left[1-\dfrac{\varepsilon a(1-z)}{(a-bz)}\right]^m}{\left[1-\dfrac{\varepsilon(1-z)}{(a-bz)}a+\dfrac{\varepsilon(1-z)a}{(a-bz)}\left(\dfrac{a-b}{a}\right)\right]^{m+1}} \quad (24)$$

Now by using the following identity [48] Eq.(25),

$$\frac{(1-t)^m}{(1-t+xt)^{m+1}} = \sum_{n=0}^{\infty} {}_2F_1(-n, m+1; 1; x) t^n \quad (25)$$

Eq.(24) can be written as

$$F(z,t) = \frac{(a-b)}{(a-bz)} \sum_{k=0}^{\infty} {}_2F_1\left(-k, m+1; 1; 1-\frac{b}{a}\right) \left(\frac{a(1-z)}{(a-bz)} \varepsilon\right)^k \quad (26)$$

By using

$$\frac{[a(1-z)]^k}{(a-bz)^{k+1}} = \frac{1}{a} \frac{(1-z)^k}{\left[1-z\left(1-\dfrac{a-b}{a}\right)\right]^{k+1}} = \frac{1}{a} \sum_{n=0}^{\infty} {}_2F_1\left(-n, k+1; 1; 1-\frac{b}{a}\right)(z)^n \quad (27)$$

Eq. (26) can be written in terms of the hypergeometric function

$$F(z,t) = \frac{(a-b)}{a} \sum_{n=0}^{\infty} \sum_{k=0}^{\infty} {}_2F_1\left(-k, m+1; 1; 1-\frac{b}{a}\right) {}_2F_1\left(-n, k+1; 1; 1-\frac{b}{a}\right)(z)^n \varepsilon^k \quad (28)$$

We now compare Eq.(20) and (28) to arrive at Eq.(29)

$$P_n(t) = \sum_{k=0}^{\infty} {}_2F_1\left(-k, m+1; 1; 1-\frac{b}{a}\right) {}_2F_1\left(-n, k+1; 1; 1-\frac{b}{a}\right) \varepsilon^k \quad (29)$$

$P_n(t)$ with the initial condition $\delta_{n,m}\delta(t)$ is the Green function $G(D^*n, D^*m, t)$. Here we denote $G(D^*n, D^*m, t)$ as $G(n,m,t)$. Therefore one can write

$$G(n,m,t) = \frac{(a-b)}{a} \sum_{k=0}^{\infty} {}_2F_1\left(-k, m+1; 1; 1-\frac{b}{a}\right) {}_2F_1\left(-n, k+1; 1; 1-\frac{b}{a}\right) \varepsilon^k \quad (30)$$

Equation (19) is expressed by the Laplace transform of *G(n,m,t)* given by

$$G(n,m,s) = \int_0^{\infty} dt\, G(n,m,t) e^{-st} \quad (31)$$

By substituting Eq.(30) into Eq.(31) and using $\varepsilon = e^{(b-a)t}$, we finally obtain

$$G(n,m,s) = \frac{(a-b)}{a} \sum_{k=0}^{\infty} {}_2F_1\left(-k, m+1; 1; 1-\frac{b}{a}\right) {}_2F_1\left(-n, k+1; 1; 1-\frac{b}{a}\right) \frac{1}{s+(a-b)k} \quad (32)$$

By using Eq.(32), we can numerically obtain the solution for Eq.(19) and the time-dependent populations.

## IV. RESULTS AND DISCUSSION

### a) Decay kinetics of excited state population relaxation

We employ the theoretical formalism detailed in **section III** and calculate the temporal relaxation of the excited state population of the donor for a model system. By using Eq.(32), we numerically obtain the solution for Eq.(19), which gives the excited state population of the donor in the Laplace plane, $P_{D^*}(s)$. Next, we perform inverse Laplace transformation to calculate the time-dependent population of the donor in the excited state, $P_{D*}(t)$ (**Figure 3**).

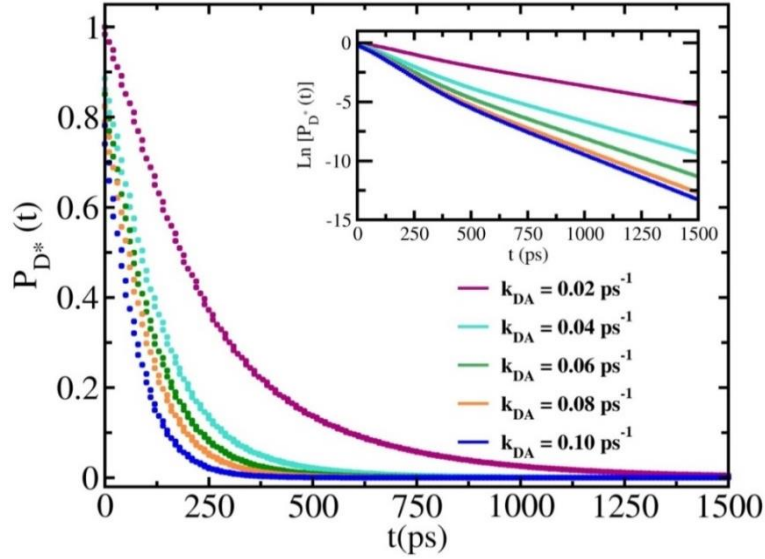

**Figure 3. The time-dependent population of the excited state of the donor is plotted for different values of intrinsic rate, $k_{DA}$ ranging from 0.02 ps$^{-1}$ to 0.10 ps$^{-1}$ at a fixed R (30 Å). We consider that the initial excitation from $S_0$ of the donor creates an excited state population in the fifth vibronic level at an arbitrary time t=0. In the inset, the semi-logarithmic plots are shown. Here vibronic energy rate ($k_{VER}$) is taken as 0.005 $ps^{-1}$ and the fluorescent emission rate is $1\,ns^{-1}$. By virtue of energy transfer, vibronic relaxation, and radiative emission, the excited state population relaxes. We observe that population relaxation is non-exponential in nature. For further study, we fit these data with a bi-exponential function [Eq. (33)]. By analyzing the fitting parameter set, the involvement of two different scales is observed. The donor acceptor energy transfer and vibronic relaxation indicate the faster and slower time scale respectively.**

The population relaxation is non-exponential in nature. We fit these data with a bi-exponential function [Eq.(33)]. The fitting parameters are given in **Table 1**.

$$F(t) = a_1 \exp\left(-\frac{t}{\tau_1}\right) + a_2 \exp\left(-\frac{t}{\tau_2}\right) \qquad (33)$$

We find that the two timescales associated with population relaxation are quite different from each other. $\tau_1$ takes the value between 10.8 and 40.2 ps, which implies that the rate of the process lies between 0.02 and 0.09 ps$^{-1}$. This faster timescale arises due to the donor-acceptor energy transfer process. On the other hand, $\tau_2$ takes the value from 125.1 to 290.2 ps which suggests that the rate of the process is in between the range of 0.003 to 0.008 ps$^{-1}$. This slower timescale arises because of vibronic relaxation. We have plotted the average lifetime of the excited donor (**Figure 4.**) by using Eq.(34).

$$\left\langle \tau_{LT}^{D} \right\rangle = a_1 \tau_1 + a_2 \tau_2 \tag{34}$$

We observe that the lifetime of the excited donor shows a monotonic and non-linear behavior with an increasing donor-acceptor energy transfer rate.

**Table 1: The population relaxation timescales are extracted by using Eq. (33). We fit the data with a bi-exponential function and obtain two different timescales. The faster timescale arises due to the donor-acceptor energy transfer process and slower timescale arises because of vibronic relaxation.**

| $k_{DA}$ (ps$^{-1}$) | $a_1$ | $\tau_1$(ps) | $a_2$ | $\tau_2$(ps) |
|---|---|---|---|---|
| 0.02 | 0.11 | 40.2 | 0.89 | 290.2 |
| 0.04 | 0.23 | 23.2 | 0.77 | 165.8 |
| 0.06 | 0.38 | 16.9 | 0.62 | 150.5 |
| 0.08 | 0.45 | 12.0 | 0.55 | 142.6 |
| 0.10 | 0.57 | 10.8 | 0.43 | 125.1 |

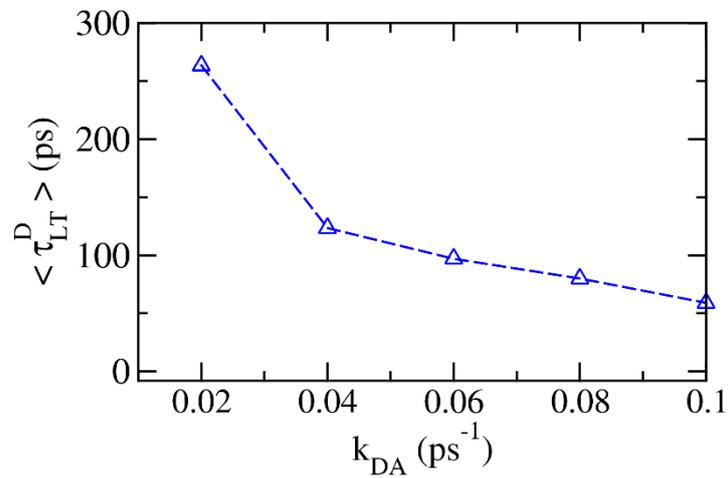

**Figure 4.** Average lifetime of excited donor is plotted against Energy transfer Rate ($k_{DA}$). Here, the vibrational energy relaxation rate $k_{VER}$ is taken to be 0.005 $ps^{-1}$, and the fluorescence emission rate $k_D$ is taken as $1\,ns^{-1}$ for all the data points. The lifetime of the excited donor exhibits a monotonic and non-linear behavior with an increasing donor-acceptor energy transfer rate.

### b) Dependence of decay on the initial population distribution

We use the model depicted in **Figure 1** to find the effect of initial population distribution. Here in the model infinite number of vibrionic levels are considered, which can be truncated with small errors. We calculate the ratio between the energy transfer rate in the presence of vibronic relaxation and the Förster rate. In order to, obtain the Förster rate, we consider a fast vibronic relaxation rate, so that energy transfer can take place from the ground vibronic level. In the present study, the fluorescent emission rate is $1\,ns^{-1}$ for the calculations.

We consider three different initial population distributions (**Figure 5b**), namely, (i) a delta function distribution, (ii) an exponential distribution, and (iii) a Gaussian distribution centred at level-3. By using Eqs. (19) and (32) we calculate the energy transfer rate in presence of vibronic relaxation and observe that the rate can be altered by a large factor (**Figure 5a**). The conventional Förster rate is calculated by considering an infinitely fast vibronic relaxation. We obtain that vibronic relaxation rate by studying the behaviour of energy transfer rate for different k$_{VER}$. We observe energy transfer rate monotonically decreases with increasing k$_{VER}$ and saturates when k$_{VER}$ > 40 ps$^{-1}$. In the present study, we consider k$_{VER}$ = 55 ps$^{-1}$ to calculate the Förster rate. To capture the non-equilibrium energy transfer scenario, slow vibronic relaxation is considered. In the presence of finite vibronic relaxation rates, the population relaxes slowly to the lower vibronic states; those states also take part in the energy transfer process. As a result, the maximum increase in the energy transfer rate (k) is observed for the distributions where a greater number of channels are associated with energy transfer. Hence, the scaled energy transfer rate ($k/k_F$) for different distributions follows the trend: Gaussian distribution (iii) > exponential distribution (ii) > delta function distribution (i)

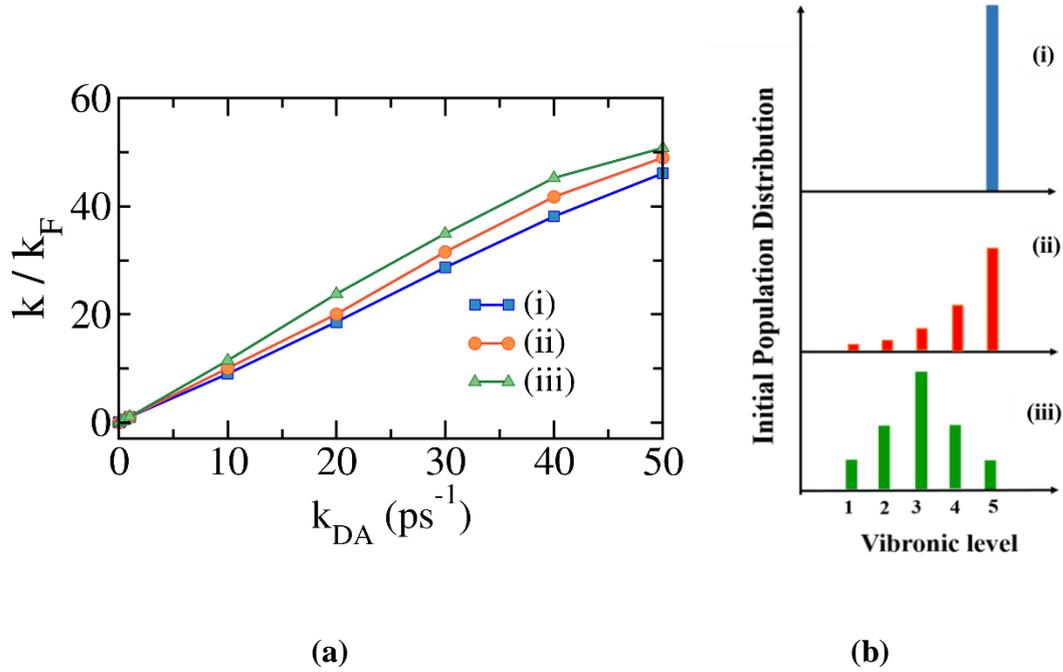

(a)                  (b)

**Figure 5. a) Scaled energy transfer rate $\left(k/k_F\right)$ is plotted against different donor to acceptor energy transfer rate $(k_{DA})$ for three different types of initial population distributions. To calculate the Förster rate ($k_F$), we consider fast vibronic relaxations. The rate is found to be monotonically decreasing with increasing $k_{VER}$, and saturates when $k_{VER} > 40$ ps$^{-1}$. We consider $k_{VER} = 55$ ps$^{-1}$ to obtain the Förster rate ($k_F$). To obtain the rates shown in Fig.5a, $k_{VER}$ is taken to be $1\ ps^{-1}$. In the high $k_{DA}$ limit, $\left(k/k_F\right)$ for different distributions follows the trend: Gaussian distribution (iii) > exponential distribution (ii) > delta function distribution (i). (b) In the right panel, the initial (t=0) population distributions are represented-i) a delta function distribution where only level-5 is populated at t=0. It's function form as $f(u') = \delta(u'-5)$, (ii) an exponential distribution from level 5 to level 1 whose functional form is considered as $f(u') = A\exp(u')$ and (iii) a Gaussian distribution centred at level-3 $f(u') = \left(1/\sqrt{2\pi\sigma^2}\right)\exp\left[-(u'-3)^2/2\sigma^2\right]$. Here $u'$ denotes vibronic level, which can take values from 1 to 5 and f($u'$) indicates the fraction of population at $u'^{th}$ vibronic level of excited donor it must must satisfy the condition $\sum_{u'=1}^{5} f(u') = 1$.**

     The altered energy transfer rate (ETR) has far reaching consequences. As energy transfer rate becomes higher than the one predicted by the Förster theory ($k_F$), the calculated value of the separation (R) between D and A would be largely overestimated. That is, the use of Förster's theory would result in a much higher value of R than the actual scenario.

     The state dependence of the initial population can be probed by varying the excitation wavelength. In the barrier-less chemical reactions, such excitation wavelength dependence can indeed play an important role in unraveling the details of its mechanism.[4] We further

study the distance dependence of the energy transfer rate for different initial population distributions. We have chosen the donor acceptor energy transfer rate ($k_{DA}$) from the earlier calculation, and Förster characteristic radius ($R_0$) is taken 60 Å. In order to obtain the donor acceptor separation distance, we use the Förster expression.

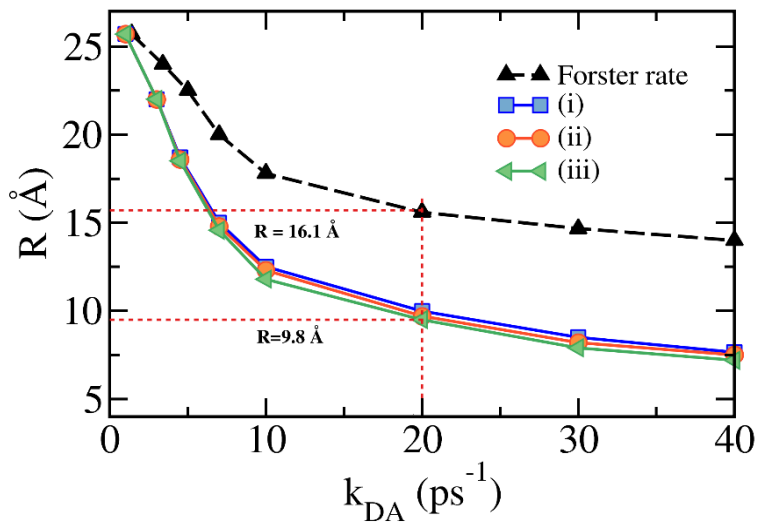

Figure 6: Distance dependence of the energy transfer rate for different initial population distributions is shown in this plot. The initial population distributions are as shown in the Figure 5.(b) [(i) a delta function distribution where only level-5 is populated at t=0, (ii) an exponential distribution from level 5 to level 1, and (iii) a Gaussian distribution centred at level-3]. We calculate the energy transfer rate (k) in the presence of slow vibronic relaxations. While calculating the Förster rate ($k_F$), we consider fast vibronic relaxation so that energy transfer can take place from the ground vibronic level. Donor acceptor separation distance (R) is calculated for both k and $k_F$ [here $R_0$ is considered to be 60Å (which can be estimated experimentally)]. The black line indicates the results obtained from the Förster rate expression. In presence of slow vibronic relaxation, for different initial population distribution, instantaneous energy transfer will occur from the intermediate vibronic level; as a result, the energy transfer rate (k) becomes higher than the one predicted by the Förster theory ($k_F$). Consequently, Förster expression overestimates the distance (R).

Here we consider the fundamental intrinsic rate dependence, and we observe that the distance predicted from the Förster theory is much higher (Black line) than that calculated distance. It is observed that the distance obtained from the Förster theory can be 40-50% higher than that of the calculated distance.

Förster theory captures a steady-state picture and it considers that the excited donor relaxed to the ground vibronic level followed by the donor acceptor energy transfer process takes place. But before reaching the steady state, donor states undergo vibronic relaxation. In the regime where excitation energy transfer is very fast compared to vibronic relaxation, the intermediate states play a key role. In that scenario, instantaneous energy transfer can take place from the non-equilibrium donor vibronic states. Due to this rapid energy transfer from the intermediate states, alteration of distance is observed.

## V. CONCLUSIONS

In order to have a numerical representation of the system in terms of a simple model, it is necessary to build a mathematical framework from which one can calculate different physical properties. Green's functions can be a powerful tool in this context. In the present work, we use the Green function formalism to study non equilibrium FRET. The process of FRET is investigated by employing a simple classical model. In order to understand the effects of the inclusion of non-equilibrium effects, we analytically treat the model system that consists of excited state vibronic manifolds for both donor and acceptor. To obtain a closed-form solution, we assume the transfer rates to be equal to each of the vibronic levels. We derive exact expressions for the temporal evolution of the excited state population for certain limiting cases.

Through our generalized Green's function based formalism on can study the decay kinetics of the excited state population relaxation and capture the non-exponential scenario (**Figure 3**). We have studied effect of the initial excited state population distribution on the energy transfer rate (**Figure 5**) for certain limiting cases. We find that our Green's function based formalism can produce a generalized analytical form of population distribution. From that result further, we have explored the distance dependence of the rate. For certain conditions, EET (Excitation energy transfer) can occur from the intermediate vibronic level, and the separation distance (R) can get affected due to the instantaneous energy transfer process. (**Figure 6**).

The main outcome of the present analysis is that the actual rate may deviate significantly from the Förster expression, the latter may predict a wrong separation distance between the donor and the acceptor. Thus the present study suggests that non-equilibrium effects, can render the estimate of the distance, R, between D and A, unreliable. A considerably enhanced rate when used in Forster expression can thus lead to a smaller estimate of R than the actual scenario. Similarly, the Förster expression can lead to a larger than the actual value of R due to breakdown of the point dipole approximation, as was shown by Wong *et al.*[6]

This formulation possesses some limitations due to the following underlying assumptions

i. The energy transfer rate from each donor level to the corresponding acceptor level is equal.

ii. Here, the back energy transfer is ignored for simplicity and for better comparison to Förster energy transfer rate expression. Also, it implies fast energy relaxation in the acceptor vibronic manifold.

However, this formalism can produce a generalized analytical form of population distribution. By using this generalized solution, one can study the effect of initial population distribution on the excitation energy transfer rate.

Förster theory describes energy transfer mechanism via incoherent hopping mechanism, which still could be applied to energy transfer in the photosynthetic complex where chromophores are widely separated, especially marine algae. The classical rate equation approach based on FRET also could be applicable in exciton diffusion[49] and could be applied in estimating efficiency in energy transfer up to a certain extent. [50,51]

A theoretical study has greater implications in understanding the energy transfer mechanism and points to developments beyond the Förster theory. We incorporate real scenarios but neglect the spatial diffusion of donor and acceptor. We have plans to incorporate it into our model and extend this approach to treat real molecules in the future.

## ACKNOWLEDGMENTS

B.B. acknowledges DST-JSPS for financial support for a visit of K.S. B.B. also thanks the Department of Science and Technology (DST, India) and Sir J.C. Bose fellowship for providing partial financial support for this work. Sangita M. acknowledges IISc, Bangalore for providing a research fellowship. Sayantan M. thanks IISc for providing a research fellowship. Finally, the authors gratefully acknowledge the SERC department at IISc for providing Mathematica license. We thank Dr. Rajesh Dutta and Dr. Sarmistha Sarkar for several important discussions and suggestions.